\documentclass[11pt]{article}
\begin{document}
\bibliographystyle{plain}
\title{\bf Issues, Concepts, and Challenges in Spintronics}
\author{ {\bf S. Das Sarma, Jaroslav Fabian, Xuedong Hu, and Igor  \v{Z}uti\'c}
\vspace*{0.3cm}
\\
\it Department of Physics, University of Maryland at College
Park \\
\it College Park, Maryland 20742-4111, USA}
\date{}

\maketitle
\thispagestyle{empty}
\small
\subsubsection*{\small ABSTRACT:}
\vspace*{-0.25cm}
We review from a theoretical perspective the emerging field of
spintronics where active control of spin transport and dynamics in
electronic materials may provide novel device application possibilities.
In particular, we discuss the quantum mechanical principles underlying
spintronics applications, emphasizing the formidable challenges involving
spin decoherence and spin injection facing any eventual device fabrication.
We provide a critical assessment of the current status of the field
with special attention to possible device applications.

\subsubsection*{\small INTRODUCTION}
\vspace*{-0.25cm}
\indent
Spintronics (sometimes also referred to as 'magneto-electronics' although
we prefer the 'spintronics' terminology because a magnetic field or the
presence of a magnetic material is not necessarily essential for
manipulating spins) is the emerging field \cite{prinz98} of active control 
of carrier spin
dynamics and transport in electronic materials (particularly, but not
necessarily limited to, semiconductors).  In some sense, existing
technologies such as GMR-based memory devices and spin valves are
elementary spintronic applications where the role of spin, however, is
passive in dictating the size of the resistance (or tunneling current)
depending on the spin direction controlled by local magnetic fields.  
Spintronics is projected to go beyond passive spin
devices, and introduce applications (and possibly whole new technologies)
based on the active control of spin dynamics.  Such active control of spin
dynamics is envisioned to lead to novel quantum-mechanical enabling
technologies such as spin transistors, spin filters and modulators, new
memory devices, and perhaps eventually quantum information processing and
quantum computation.  The possibility of monolithic integration on a single
device of magnetic, optical, and electronic applications, where magnetic
field and polarized light control spin dynamics, is an exciting new
spintronic prospect for creating novel magneto-electro-optical technology.
The two important physical principles underlying the
current interest in spintronics are the inherent quantum mechanical nature
of spin as a dynamical variable (leading to the possibility of novel
spintronic quantum devices not feasible within the present-day charge-based
electronics) and the inherently long relaxation or coherence time
associated with spin states (compared with the ordinary momentum states).
The fact that carrier spin in semiconductors can be easily manipulated
noninvasively by using local magnetic fields, by applying external electric
fields through controlled gates, and even by shining polarized light is an
important impetus for developing spintronics applications.

In spite of the great current interest in the basic principles and concepts
of spintronics a large number of obstacles need to be overcome before one
can manufacture spintronics applications. 
For example, a basic spintronics transport requirement is to produce and
sustain large spin-polarized currents in electronic materials
(semiconductors) for long times.  This has not yet been accomplished.  In
fact, it has turned out to be problematic to introduce spin-polarized
carriers in any significant amount into semiconductor materials.
Similarly, for quantum computation one requires significant and precisely
controllable spin entanglement as well as single spin (i. e., a single Bohr
magneton) manipulation using local magnetic fields.  Currently there is no
good idea about how to accomplish this.  It is clear that a great deal of
basic fundamental physics research will be needed before spintronics
applications become a reality.

In this paper we highlight and summarize a few examples of spintronics
research with the emphasis on understanding principles and operations with
future device potential.  We 
concentrate on the elementary aspects of spintronics which must be
understood and developed before any possible applications can be discussed.
These aspects are creating, maintaining, manipulating,
and measuring spin currents in semiconductors (and related electronic
materials), spin entanglement in semiconductor quantum dots in the context
of quantum computation, and spin relaxation.  The examples are drawn from
our own theoretical research on spintronics, and we refer the reader to our
existing publications for the details.

\subsubsection*{\small SPIN RELAXATION AND DECOHERENCE} 
\vspace*{-0.25cm}
The great promise of spintronic technology is based upon the fundamental ability
of electron spins in electronic materials to preserve
coherence for relatively long times. A typical electron ``remembers'' its initial
spin orientation for a nanosecond. This time scale is indeed long when 
compared with the typical times--femtoseconds--for electron momentum relaxation.
Perhaps a more revealing quantity than spin lifetime (which is usually called spin 
relaxation time $T_1$ or spin decoherence time $T_2$, depending on the context of 
the experiment) is the spin diffusion length $L_S$ which measures how far 
electrons diffuse in a solid without losing spin coherence. The
important fact that $L_S$ is typically a micrometer makes 
spintronics a viable option for future micro- and 
nanoelectronics; any information encoded in electron spins will spread
undisturbed throughout the device. Clearly, the longer the spin lifetime, 
the better and more reliable will be the spintronic devices. The study
of spin relaxation is thus of great importance for spin-based technology (we 
reviewed the current understanding of spin relaxation processes in electronic 
systems in Ref. \cite{fabian99}).  

Initial measurements of spin lifetimes were conducted in metals like Na or Li by
conduction electron spin resonance (CESR) technique \cite{feher55}. 
The most important 
outcome of these experiments concerned the magnitude of $T_1$ (nanoseconds)
and its temperature behavior: $T_1$ is constant at low temperatures (below, say,
50 K) and is increasing linearly with increasing temperature at elevated 
temperatures (above, say, 200 K). These two observations helped to shape the 
theoretical understanding of the processes behind spin relaxation in metals.
It is now generally accepted that electron spins in (nonmagnetic) metals decay
by scattering off impurities (at low temperatures where $T_1$ is constant) and 
phonons (at higher temperatures where $T_1$ grows linearly with increasing 
temperature). The spin-flip probability of such processes is finite because
of the finite spin-orbit interaction induced by either host ions or impurities
(this is the so called Elliott-Yafet mechanism of spin relaxation
\cite{elliott54}). We have
recently performed the first realistic calculation of $T_1$ in a metal (aluminum)
\cite{fabian98}.
Our calculation not only provides the first direct proof of the validity of the
Elliott-Yafet mechanism, but also shows that by engineering the band structure
of metals (or semiconductors) it is possible to tailor 
spin relaxation (e.g., $T_1$ can be changed by orders of magnitude by doping,
straining, alloying, or changing dimensionality).   

An important development came with the discovery of spin injection
by Johnson and Silsbee \cite{johnson85}. In the original experiment 
spin-polarized electrons were injected from a ferromagnetic electrode (permalloy)
into a nonmagnetic metal (aluminum), and the spin diffusion length was monitored. 
This method of measuring $L_S$ (and thus spin lifetime) has
a great potential since, unlike CESR, spin injection does not need an applied magnetic field
which, in some cases, radically affects spin relaxation processes. In addition to providing
a useful method for measuring spin relaxation, the Johnson-Silsbee spin injection
experiment brought about a whole new field of electronics: spintronics. Indeed,
spin injection is the most natural way to integrate spin dynamics with electronic
transport in electronic devices. There is no need for magnetic field or radiation
to excite spin-polarized electrons. One only needs ferromagnetic electrodes. The 
last truly fundamental obstacle in the progress towards integrating the new 
spintronic with the traditional semiconductor technology has been recently overcome
with the discovery of spin injection into a 
semiconductor \cite{hammar99,semiconductor}.

In addition to be able to create the population of spin-polarized carriers, we 
also need a way to monitor and control the dynamics of spin processes in 
electronic materials. This quest has been pioneered by Kikkawa and 
Awschalom \cite{awschalom99}. In a typical experiment spin-polarized electrons 
in a semiconductor like GaAs are excited by a circularly-polarized light and 
then the electrons' spin evolution is monitored at small (picosecond) time 
intervals. Several new exciting results came from such experiments on
semiconductors \cite{kikkawa97}: a dramatic (two orders of magnitude) increase of electron spin 
lifetime with increased doping, 
unusually large (hundreds of micrometers) spin diffusion length, 
and the ability to optically control nuclear
spin polarization
(with electron spins acting as intermediaries between
light and nuclear spins).


\subsubsection*{\small SPIN-POLARIZED TRANSPORT}
\vspace*{-0.25cm}
The goal of employing both spin and charge transport
(spin-polarized transport) in potential
novel device applications imposes intrinsic limitations on their design:
they should consist of either heterostructures or inhomogeneous
materials. While  similar design constraints have been extensively
investigated and well understood in the case of pure charge transport
in conventional electronics, it is not clear how the spin degrees of freedom
will behave in transport  across interfaces in a
heterostructure or through an inhomogeneous material. 
For example, by placing a semiconductor in  contact with a nonmagnetic
metal a Schottky barrier is formed whose properties will govern
charge transport across the semiconductor/metal junction. 
Currently there is no physical understanding for the corresponding spin-dependent
Schottky barrier relevant for spin-polarized transport across interfaces.
This is an important issue as some of the proposed spintronic 
devices \cite{datta90} rely on the direct electrical spin injection from a 
ferromagnet into a semiconductor \cite{hammar99,semiconductor}. The situation
is further complicated by the possibility of spin-flip scattering at
magnetically active interfaces. These considerations 
have to be included in assessing the feasibility of various spintronic devices
because they imply that
the degree of carrier spin-polarization can be strongly modified during 
transport across semiconductor/ferromagnet interfaces.

Fabricating hybrid structures which would combine a semiconductor and a
superconductor would allow investigating some of the aforementioned features  
and determining the degree of an extrinsically induced  carrier spin
polarization in the semiconductor.
This could be realized by
using  Andreev reflection 
which governs transport properties at low applied bias. 
In this two-particle process  an
electron incident to the interface at the semiconductor side is accompanied 
by a second electron of the opposite spin. 
Both electrons are then 
transfered into the superconductor where they form a Cooper pair.
The probability (measured by, e.g., conductance) of such processes 
strongly depends on the amount of spin polarization and the spin transparency
of the interface \cite{zutic99,zutic99a}.

Material inhomogeneities can also act
favorably and be tailored to give desired effect for spin-polarized
transport.
We illustrate this
in  our proposal of the spin-polarized p-n junction \cite{zutic00}. 
Its simple realization
would consist of  shining circularly-polarized light on the p-doped side of a
usual p-n junction. This would create a spin-polarized
population of electron-hole pairs.
By considering a p-n junction shorter than the spin-diffusion
length, combined with a sharp doping profile, it is feasible to create 
enhanced magnetization in the interior of the semiconductor with the spatial 
dependence following that of the carrier concentration. Such a p-n junction 
could  be used as a building block for a novel spin transistor applications which 
would utilize both spin and charge degrees of freedom \cite{zutic00}.
   
\subsection*{\small SPIN-BASED QUANTUM COMPUTATION}
\vspace*{-0.25cm}
Among all possible spintronic devices, by far the most revolutionary is
the proposed spin-based quantum 
computer (QC) which has  the promise to vastly outperform classical
computers in certain tasks such as factoring large numbers
and searching large databases \cite{QCreview}.  In  
QCs electron or nuclear spins are used as the basic 
building blocks.  The spin-up and -down states of an electron
or a nucleus provide the quantum bit (qubit), in analogy with 
``0'' and ``1'' in a classical computer.  However, as a
quantum mechanical object, a spin can have not only up
and down states, but also arbitrary superpositions of 
these two states.  This inherent parallelism and other quantum
mechanical properties such as entanglement and unitary evolution 
are the fundamental differences between QCs and
classical computers.  

In the search for appropriate hardwares for a QC, many proposals
have been put forward \cite{QCreview}.  Here we focus on the
spin-based solid-state models \cite{HD}.  One of the first proposals \cite{LD} suggests
using quantum-dot-trapped electron spins as qubits.  Here a 
single electron is trapped in a gated horizontal GaAs quantum dot,
with pulsed local magnetic field and inter-dot gate voltage governing
the single-qubit and two-qubit operation.  Another proposal 
replaces the quantum dot electrons by donor electrons 
\cite{Vrijen}.  Here varying the gyromagnetic ratio in a compositionally
modulated SiGe alloy allows electron spin resonance for single qubit 
operations and exchange interaction for two-qubit operations.  
One important advantage of electron 
spins is their ``maneuverability'': electrons are mobile and
can be manipulated by both electric and magnetic fields.

Aside from electron-spin-based QC models, there are also 
nuclear-spin-based proposals, such as the one using nuclear spins
of phosphorus donor atoms in Si as qubits \cite{Kane}.  Here
external gates are used to tune the nuclear magnetic resonance
frequency, and donor electrons are the intermediaries between
neighboring nuclear spins, introducing two-qubit operations
through electron exchange interaction and hyperfine interaction.
The main advantage of 
nuclear spin qubits is their exceedingly long coherence time,
which allows many coherent operations.  Indeed, bulk solution
NMR is one of the most advanced QC architectures
\cite{QCreview}, even though its ensemble-average character 
prompts some researchers to question \cite{Caves} whether 
it really possesses all the quantum mechanical powers needed
for tasks such as factoring.

The major difficulties facing various QC models are achieving
precise control over unitary evolutions and maintaining
quantum mechanical coherence.  While traditional electronic
devices deal with large numbers  of electrons at a time, while in 
spin-based QCs one has to be able to precisely control spins of individual 
electrons.  Furthermore, the electron spins
need to be essentially isolated from their environment so that their
dynamics is governed by quantum mechanics.  If this isolation is
imperfect, the spins' quantum information will leak into their 
environment, and the dynamics of the spins will become irreversible
and classical, so that the QC operation will be disrupted.

Spin decoherence has many different channels such as spin interaction
with boundaries, impurities, host nuclei, even with external controls.
For example, one common approach to tune the exchange interaction
between electrons or electrons and nuclei is to use electrical gates
which are connected through a transmission line to the outside.  
External noise such as Johnson-Nyquist noise can thus cause fluctuations
in the gate voltage, which in turn cause errors in the exchange.  The
rate of this error can be as large as a few MHz \cite{HD}, which corresponds
to the limits of the currently available error correction
schemes.  Another error during exchange is caused by 
inhomogeneous magnetic fields \cite{Inhomo}.  In essence, the different
Zeeman couplings of two neighboring electrons cause mixing of the
two-electron singlet and triplet states, therefore preventing the
electron spin states from complete disentanglement for swap.  This
error is proportional to the square of the inhomogeneity \cite{Inhomo},
and can usually be corrected.  Indeed, there is an existing scheme
which can circumvent this error \cite{parallel}.  Furthermore, it has
been proposed \cite{Lidar} that one can utilize certain decoherence-free
subspace  of four quantum dots as qubits, relying completely on exchange 
for all operations and eliminating the use of any external magnetic field.
Such a scheme is more difficult to realize experimentally, but it does
provide the advantage of smaller decoherence because of fewer noise 
channels. 



\subsubsection*{ACKNOWLEDGMENT:}
\vspace{-0.25cm}
This work was supported by the US ONR, DARPA, and LPS.

\end{document}